\documentclass[english,twocolumn]{article}

\usepackage[utf8]{inputenc}				
\usepackage[big,online]{dgruyter}	
\setlocalecaption{german}{abstract}{Zusammenfassung} 

\usepackage{lmodern} 
\usepackage{microtype}
\usepackage[numbers,square,sort&compress]{natbib}

\usepackage{bm}
\usepackage{mathtools, cuted}

\usepackage{hyperref}
\hypersetup{
    colorlinks   = true, 
    urlcolor     = blue, 
    linkcolor    = black, 
    citecolor    = blue, 
    pdftitle     = {Local Bayesian Optimization for Controller
Tuning with Crash Constraints}, 
    pdfauthor    = {Alexander von Rohr, David Stenger, Dominik Scheurenberg, Sebastian Trimpe}, 
}

\usepackage{tikz}
\usetikzlibrary{arrows.meta,positioning,fit,calc}
\usetikzlibrary{shapes.geometric, arrows, positioning}
\usetikzlibrary {backgrounds}
\definecolor{RWTHBlue}{rgb}{0, 0.32941176470588235, 0.6235294117647059}

\usepackage{acronym}

\acrodef{gp}[GP]{Gaussian process}
\acrodef{bo}[BO]{Bayesian Optimization}
\acrodef{gibo}[\textsc{GIBO}]{Gradient Information with BO}
\acrodef{vdpbo}[\textsc{VDP}-\ac{bo}]{\ac{bo} with virtual data points}
\acrodef{crashgibo}[\textsc{VDP}-\ac{gibo}]{\ac{gibo} with virtual data points}
\acrodef{mpc}[MPC]{model predictive control}
\acrodef{lqi}[LQI]{linear quadratic integral}
\acrodef{doe}[DoE]{design of experiments}

\acrodef{bo_german}[BO]{Bayes’sche Optimierung}

\theoremstyle{dgthm}

\newtheorem{assumption}{Assumption}

\theoremstyle{dgdef}

\DeclareMathOperator*{\argmin}{arg\,min}
\DeclareMathOperator{\Tr}{Tr}

\newcommand{\loc}{\bm{x}}
\newcommand{\locentry}{x}
\newcommand{\Loc}{X}
\newcommand{\domain}{\mathcal{X}}\newcommand{\state}{\bm{s}}
\newcommand{\action}{\bm{a}}

\newcommand{\ie}{i\/.\/e\/.,\/~}
\newcommand{\eg}{e\/.\/g\/.,\/~}
\newcommand{\cf}{cf\/.\/~}

\usepackage{algorithm,algpseudocode}
\makeatletter
\def\plist@algorithm{Alg.\space}
\makeatother

\begin{document}

\startpage{1}

\title{Local Bayesian Optimization for Controller Tuning with Crash Constraints}
\runningtitle{Local Bayesian Optimization for Controller Tuning}

\author*[1]{Alexander von Rohr}
\author[2]{David Stenger}
\author[2]{Dominik Scheurenberg}
\author[3]{Sebastian Trimpe}
\runningauthor{A.~von Rohr et al.}
\affil[1]{\protect\raggedright 
Institute for Data Science in Mechanical Engineering, RWTH Aachen University, Germany, e-mail: vonrohr@dsme.rwth-aachen.de}
\affil[2]{\protect\raggedright 
Institute of Automatic Control, RWTH Aachen University, Germany, e-mail: \{d.stenger,d.scheurenberg\}@irt.rwth-aachen.de}
\affil[3]{\protect\raggedright 
Institute for Data Science in Mechanical Engineering, RWTH Aachen University, Germany, e-mail: trimpe@dsme.rwth-aachen.de}
	

\abstract{
Controller tuning is crucial for closed-loop performance but often involves manual adjustments. Although \ac{bo} has been established as a data-efficient method for automated tuning, applying it to large and high-dimensional search spaces remains challenging. We extend a recently proposed local variant of \ac{bo} to include crash constraints, where the controller can only be successfully evaluated in an a-priori unknown feasible region. We demonstrate the efficiency of the proposed method through simulations and hardware experiments. Our findings showcase the potential of local \ac{bo} to enhance controller performance and reduce the time and resources necessary for tuning.
}

\maketitle

\acresetall
\section{Introduction} 
Most control algorithms involve user-defined parameters that determine the closed-loop behavior.
Examples include controller gains for PID-controllers and stage and terminal costs in \ac{mpc}.
Inadequate choices for these parameters often lead to performance issues \cite{jelali2006overview}.
Controller tuning is the process of adjusting parameters to meet specified performance requirements for a given control task. Evaluation of the performance requires running experiments in either simulation or on hardware.
Although analytical solutions for optimal parameters exist in some cases, for instance, for the linear quadratic regulator (LQR) and \ac{lqi} control, practical applications often require adjustments of the weighting matrices to ensure the closed-loop satisfies performance requirements that are not captured in those cost functions or to counteract modeling inaccuracies.

\begin{figure}
    \centering
    \tikzset{
      frameblack/.style={
        rectangle, draw,
        text width=10em, text centered, text=white,
        minimum height=5em,fill=black!40,
        rounded corners,
      },
      frameblue/.style={
        rectangle, draw,
        text width=8em, text centered,
        minimum height=2em,fill=RWTHBlue!10,
        rounded corners,
      },
      line/.style={
        draw, -{Latex},rounded corners=3mm,
      },
    }       
    \begin{tikzpicture}[font=\large,very thick,node distance = 4cm]

        \node [frameblue] (environment) {plant};

        \node [frameblue, below=0.4 of environment] (agent) {controller $\pi_{\loc}$};
        
        \draw[line] (agent.west) -- ++ (-1,0) |- (environment) node[right,pos=0.25,align=left] {\small input};
    
        \draw[line] (environment.east) -- ++ (1,0) |- (agent.east) node[left, pos=0.25, align=right] {\small state};

        \begin{scope}[on background layer]
        \coordinate[left=2.5 of environment.north] (upper_left);
        \coordinate[right=2.5 of agent.south] (lower_right);
        
        \node[frameblack,style=dashed, fit=(agent) (environment) (upper_left) (lower_right)] (system) {};
        \end{scope}

        \node [frameblue,text width=10em,below=0.4 of system] (optimizer) {Bayesian \\ optimization
        \small $\loc^* = \argmin_{\loc \in \domain} f(\loc)$
        };

        \draw[line] (optimizer.west) -- ++ (-1.5,0) node[below, pos=0.5, align=left] {\small $\loc \in \domain$} |- (system.west) ;

        \draw[line,style={Latex}-] (optimizer.east) -- ++ (1.5,0) node[below, pos=0.5, align=left] {\small $f(\loc) + \epsilon$} node[above, pos=0.5, align=left] {\small if $\loc \in \domain_S$} |- (system.east) ;

\end{tikzpicture}
    \caption{
    The controller tuning process with \ac{bo}. The objective $f$ is evaluated in closed-loop. The controller $\pi_{\loc}$ has tuning parameters $\loc \in \domain$ and \ac{bo} searches for the optimal parameterization. No function value is available if an experiment crashes, $\loc \not\in \domain_S \subseteq \domain$. 
    }
    \label{fig:controller-tuning}
\end{figure}
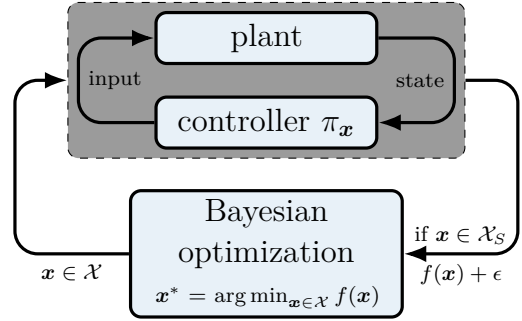

Automation presents a promising solution to improve control performance during commissioning and in response to changes in operating conditions.
Automated controller tuning aims to identify effective controllers by utilizing prior knowledge about the plant and data collected during its operation. An emerging approach in controller tuning is \ac{bo}, which is particularly well-suited for this purpose due to its data-efficiency \cite{chatzilygeroudis2020survey,paulson2023tutorial}.
The controller tuning loop with \ac{bo} is illustrated in Fig.~\ref{fig:controller-tuning}. 

A well-known limitation of \ac{bo} in a practical controller tuning setting is its dependence on the dimension of the search domain. Controller tuning involves evaluating the closed-loop performance through experiments, each of which can take several minutes. Consequently, conducting more than a few hundred evaluations can be impractical. Therefore, an automated tuning method should reliably identify effective controllers within a few evaluations. This requirement restricts \ac{bo}'s applicability to low-dimensional problems and small search domains.

Practical controller tuning can also introduce an additional layer of complexity in the form of crash constraints \cite{bachoc2020gaussian}. If control performance is evaluated by observing the closed-loop system, it may exhibit unsafe or undesired behavior, which requires terminating the experiment early. This often means that successful and crashed evaluations cannot be meaningfully expressed in the same metric. Crash constraints are common when tuning controllers for complex systems, such as in robotics and have already been present in tuning problems considered in the first works on \ac{bo} for controller tuning (\eg\cite{Marco.2016, Calandra.2016}). The ubiquity of crash constraints motivated specialized learning methods such as \citet{Marco.2021}.

In previous work, we proposed \ac{gibo} \cite{mueller2021local}, a local variant of \ac{bo}. Local search begins with an initial parameterization, and \ac{gibo} uses a small number of evaluations to learn a descent direction to update the parameters in an improvement step.
Empirical studies show that \ac{gibo} and its variants \cite{nguyen2022local,wu2023behavior} are more data-efficient and outperform global \ac{bo} on synthetic benchmarks.
However, the \ac{gibo} algorithm has not yet been applied to practical control problems. In this article, we revisit \ac{gibo} and investigate its applicability for practical controller tuning under crash constraints through simulation and hardware experiments.

The benefits of local search for controller tuning include higher data-efficiency for large and high-dimensional problems, continuous improvement, and local exploration.
As an additional benefit, control engineers and other algorithm users may find local search more intuitive due to smaller updates, making these algorithms easier to understand and deploy.
Nevertheless, local optimization is sensitive to the initial parameterization and may converge to sub-optimal local minima. 
Fortunately, prior work indicates that controller tuning problems frequently have a unique minimum \cite{stenger2022benchmark}.

\subsection{Problem Statement}\label{ssec:problem}

The controller tuning problem is defined as optimizing an objective function which maps control algorithm parameters to the performance of the closed-loop system
\begin{align}\label{eq:opt}
    \loc^* = \argmin_{\loc \in \domain_S} f(\loc),
\end{align}
where $\domain_S \subseteq \domain \subset \mathbb{R}^d$ is the feasible region of the search space $\domain$ and $d$ is its dimensionality. The crash region is denoted as $\domain_C = \domain \setminus \domain_S$.
The objective $f$ is a costly black-box function; that is, evaluations are required to obtain its function value, and these evaluations are resource-intensive. 
Note that the feasible region $\domain_S$ and, therefore, the crash region $\domain_C$ are unknown, and $f$ is undefined outside of it. Evaluating $f$ outside of $\domain_S$ remains costly but will not yield an objective value.
This problem formulation requires that there is an experimental procedure in place to evaluate performance and recover from a crashed evaluation. These procedures may be fully automated but can also require human interventions. 

We assume we can collect data of the form $(\loc, y)$ for any $\loc \in \domain_S$ with 
\begin{equation*}\label{eq:obs}
    y = f(\loc) + \epsilon,
\end{equation*}
where $\epsilon \sim \mathcal{N}(0,\sigma^2_n)$ is independent and identically distributed Gaussian noise.
We denote $\mathcal{D} = (X, \bm{y})$ as the dataset with $|\mathcal{D}| \coloneqq N$ and
\begin{equation*}
    X \coloneqq \begin{bmatrix}
        x_1, \\
        \vdots \\
        x_N
    \end{bmatrix} \; 
    \bm{y} \coloneqq \begin{bmatrix}
        y_1, \\ 
        \vdots \\
        y_N
    \end{bmatrix}.
\end{equation*}
No direct assumptions are made regarding the controlled system or control algorithm. However, we assume the objective function is a sample from a mean-square differentiable \ac{gp}. This assumption enables us to learn a local gradient and determine a search direction that enhances closed-loop behavior. It is worth noting that this type of regularity assumption is a standard practice in \ac{bo} \cite{srinivas2010gaussian}. For practical purposes, it is common to select compact and convex sets as the search domain $\domain$. 
\begin{assumption}
\label{ass:gp}
The performance function $f$ is a sample from a Gaussian process with $p(f) = \mathcal{GP} (f; \mu, k)$, whose mean function $\mu : \domain \to \mathbb{R}$ is at least once differentiable and whose covariance function $k : \domain \times \domain \to \mathbb{R}$ is at least twice differentiable. 
\end{assumption}

\subsection{Contributions}
We propose \ac{crashgibo}, a general and data-efficient optimization algorithm for controller tuning under crash constraints~\eqref{eq:opt}. The proposed method is based on prior work on local \ac{bo} \cite{mueller2021local}, which learns the gradient of the control objective with respect to the tuning parameters from noisy closed-loop performance evaluations. Crash constraints are addressed by introducing virtual data points \cite{stenger2022benchmark}, which guide the optimization away from the infeasible region. The proposed method is evaluated on a simulated coupled tank system with popular control algorithms, namely PI control, \ac{lqi} and \ac{mpc}. Additionally, we validate our method by tuning a PI controller on hardware.

\section{Related Work}

This section is an overview of different controller tuning applications using BO. BO treats the tuning task as a black-box optimization problem. It is not restricted to a special system or objective function class. Therefore, it has been used to automatically optimize the parameters of different controller structures such as LQR \cite{Calandra.2016},
\ac{mpc} \cite{Andersson.16.05.201621.05.2016}, and PID \cite{Chen.2019}.
In addition to controllers, BO has been applied to other control engineering algorithms such as Kalman filter \cite{Chen.71020187132018} and fault diagnosis \cite{MARZAT201112904}.
BO can be applied to complex hierarchical controller structures and the interaction between controllers and filters \cite{MohammadKhosravi., Stenger.2022}.

The problem statement in \eqref{eq:opt} is a single-objective formulation with crash constraints. However, in practical controller tuning problems, more complex formulations may be required to capture the tuning task fully. With its various extensions, \ac{bo} offers a versatile toolkit to address those issues. In contextual optimization \cite{fiducioso2019safe}, parameters are optimized as a function of an operating condition. Optimization with unknown constraints (\eg\cite{khosravi2023safety}) restricts the set of feasible solutions. In contrast to the setting herein, safe BO tries to stay within those bounds also \emph{during} optimization \cite{berkenkamp2021bayesian}.
Pareto optimization simultaneously considers multiple objectives \cite{math11020465,makrygiorgos2022performance}.
Related topics include robust optimization \cite{Frohlich.2020, paulson2022adversarially},
preference-based tuning \cite{ZhuMengjia.2021}, 
and time-varying optimization problems \cite{brunzema2022controller}.     

Approaches to address crash constraints in controller tuning include assigning a fixed penalty (\eg\cite{Marco.2016}) or using data obtained before the crash (\eg\cite{Calandra.2016}). However, it may require substantial domain knowledge
to design the penalty.
A probabilistic classifier in combination with constrained BO (\eg\cite{STENGER2020, bachoc2020gaussian}) can also be used to address the issue. However, this may result in tedious tuning of the additional hyperparameters of the classifier.
\citet{Marco.2021} propose a combined \ac{gp} model for constrained optimization with crash constraint and apply it to a controller tuning task on a quadroped robot. This approach requires modifying the acquisition function to incorporate the separate model of the constraints.  

Herein, we address crash constraints using \ac{vdpbo}. The upside of \ac{vdpbo} is that it only modifies the \ac{gp} modeling step of \ac{bo} by introducing virtual observation. Therefore, the acquisition function step remains unchanged, and \ac{vdpbo} can be easily incorporated with different \ac{bo} flavors. \ac{vdpbo} was introduced for single-objective optimization in \cite{stenger2022benchmark}, and applied to constrained \cite{Stenger.2022} and multi-objective optimization \cite{math11020465}.

\section{Preliminaries}
This section introduces Gaussian Processes (\acp{gp}) and their derivatives, along with a strategy for minimizing the posterior variance of gradient estimates. For an introduction to \acp{gp} and \ac{bo}, we refer to \citet{garnett2023bayesian}. 

\subsection{Gaussian Process Derivatives}

\begin{figure*}[t]
    \centering
    \input{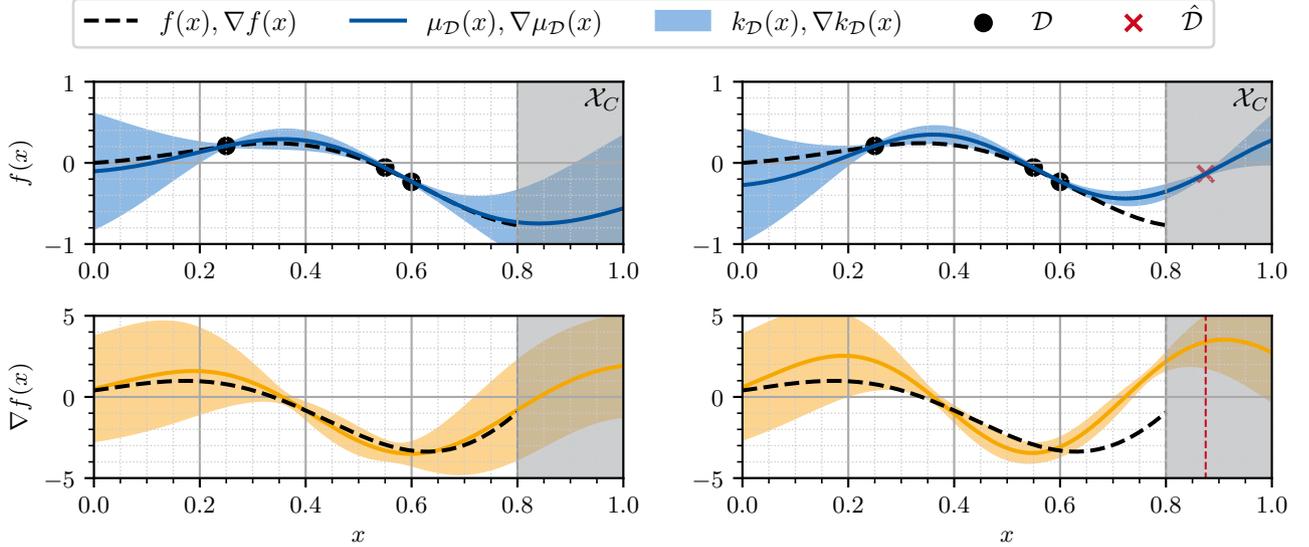}
    \caption{\textbf{Left:} A Gaussian process posterior (top) and its derivative (bottom). \textbf{Right:} The posterior with an additional virtual observation in the crash region $\domain_C$. 
    The crashed evaluation (red cross) cannot be evaluated, and a virtual observation is added instead. In this example, the virtual data point modifies the posterior such that the minimum of the posterior is not inside the infeasible region $\domain_C$, and the gradient points away from it.}
    \label{fig:gp}
\end{figure*}

We utilize a \ac{gp} model to model the expectation and the uncertainty of the objective's gradient. This model guides the optimization procedure to improve the closed-loop performance quickly.
Given the prior from Assumption~\ref{ass:gp} and a dataset of closed-loop performance observations $\mathcal{D}$, their joint distribution is
\begin{equation*}
    p(f, y) = \mathcal{GP}\left( \begin{bmatrix}
        f \\ \bm{y}
    \end{bmatrix};
    \begin{bmatrix}
        \mu \\ \bm{m}
    \end{bmatrix},
    \begin{bmatrix}
        k & \kappa^\top \\ \kappa & C 
    \end{bmatrix}
    \right),
\end{equation*}
where $\bm{m} = \mu(X)$, $C = k(\Loc, \Loc) + \sigma^2_n I$ , and $\kappa = k(\cdot, X)$.
Mean and covariances are given by $\mu$ and $k$.
The posterior distribution at location $\loc_*$ is (\cf\cite{rasmussen2006gaussian})
\begin{equation*}\label{eq:gp_post}
    p(f(\loc_*) \mid \mathcal{D}) = \mathcal{N}\left( f(\loc_*);
    \mu_\mathcal{D}(\loc_*), k_\mathcal{D}(\loc_*)
    \right),
\end{equation*}
where
\begin{align*}
    \mu_\mathcal{D}(\loc_*) &= \mu(\loc_*) + \kappa(\loc_*)^\top C^{-1} (\bm{y} - \bm{m}) \\
    k_\mathcal{D}(\loc_*) &= k(\loc_*, \loc_*) - \kappa(\loc_*)^\top C^{-1} \kappa(\loc_*). 
\end{align*}
Analogously, the joint distribution between observations and the functions derivative is 
\begin{equation*}
    p(\nabla f, y) = \mathcal{GP}\left( \begin{bmatrix}
        \nabla f \\ \bm{y}
    \end{bmatrix};
    \begin{bmatrix}
        \nabla \mu \\ \bm{m}
    \end{bmatrix},
    \begin{bmatrix}
        \nabla k \nabla^\top &  (\nabla\kappa)^\top\\ \nabla \kappa & C 
    \end{bmatrix}
    \right),
\end{equation*}
where $\nabla$ is the differential operator and $\nabla$ placed behind $k$ takes the derivative w.r.t. the second input.
The posterior of the derivative at $\loc_*$ is
\begin{equation*}\label{eq:dgp_post}
    p(\nabla f(\loc_*) \mid \mathcal{D}) = \mathcal{N}\left( \nabla f(\loc_*);
    \nabla\mu_\mathcal{D}(\loc_*), \nabla k_\mathcal{D}(\loc_*)
    \right),
\end{equation*}
where
\begin{align*}
    \nabla\mu_\mathcal{D}(\loc_*) &= \nabla\mu(\loc_*) + \nabla\kappa(\loc_*)^\top C^{-1} (\bm{y} - \bm{m}) \\
    \nabla k_\mathcal{D}(\loc_*) &= \nabla k(\loc_*, \loc_*) \nabla^\top - \nabla\kappa(\loc_*)^\top C^{-1} \nabla\kappa(\loc_*). 
\end{align*}
This \ac{gp} is the posterior distribution over the gradient based on zeroth-order information, meaning observation of the objective function. It is also possible to incorporate gradient observation if available. 
For a depiction of a one-dimensional \ac{gp} and its derivative, see Fig.~\ref{fig:gp}.
The posterior over derivatives is a vector-valued \ac{gp} for $d > 1$.

\subsection{Gradient Uncertainty}\label{ssec:gi}

In our method, the objective of an experiment is finding a descent direction by minimizing the gradient uncertainty for a given parameterization $\loc_*$. We achieve this by an optimal \ac{doe}.
We define an extended dataset $\mathcal{D}' = \mathcal{D} \cup \{(\Loc',y')\}$ which includes future observations at $\Loc'$ and unknown value $y'$. 
We denote the total variance as the sum of the eigenvalues of the covariance matrix at $\loc_*$, which is its trace $\Tr(\nabla k_\mathcal{D}(\loc_*))$.
The posterior total variance at $\loc_*$ after $b$ additional future observations at $\Loc' = \domain \times \mathbb{R}^{d \times b}$ is (\cf\cite{mueller2021local})
\begin{equation}\label{eq:gi}
    \begin{aligned}
        &\alpha_\text{TV}(x_*, X') =  \Tr \left( \nabla k_\mathcal{D'}(\loc_*) \right) \\ 
    \end{aligned}
\end{equation}
where $\nabla k_\mathcal{D'}(\loc_*)$ is posterior variance based on the extended dataset $\mathcal{D}'$.
The total variance in \eqref{eq:gi} does not depend on the unknown future observation $y'$ and only on the location of this observation, allowing for an analytic expression of \eqref{eq:gi}. Consequently, a \ac{doe} can be computed by minimizing \eqref{eq:gi} over the controller parameters to be evaluated in the next batch $\Loc_{\textrm{DoE}}$
\begin{equation}\label{eq:gibo_doe}
    \Loc_{\textrm{DoE}} = \argmin_{X' \in \domain \times \mathbb{R}^{d \times b}} \alpha_{TV}(x_*, X').
\end{equation}
Minimizing the total variance is equivalent to minimizing the quadratic distance of samples from the gradient distribution, which also minimizes the worst-case gradient estimation error \cite{wu2023behavior}.
As an alternative to a \ac{doe} based on the total variance \citet{nguyen2022local} propose to maximize the probability of descent.

\section{Proposed Method}

The method proposed in this article is a local \ac{bo} approach designed to deal with the crash-constrained optimization problems often encountered in controller tuning.
It is an extension of the \ac{gibo} method proposed in \citet{mueller2021local} and incorporates virtual data points for crashed evaluations introduced by \citet{stenger2022benchmark}. The resulting optimization algorithm is summarized in Algorithm~\ref{algo:cgibo}.

\subsection{Virtual Data Points for Crashed Evaluations}\label{ssec:vdata}

As stated in Section~\ref{ssec:problem}, the feasible domain $\domain_S$ is unknown.
The optimization algorithm might try to evaluate $f$ outside of this region, crash, and not receive a value for $y$. Nevertheless, we must incorporate this failed evaluation in the \ac{gp} model.
A naive approach is to replace the function evaluation with a virtual observation of a fixed penalty for violating the crash constraint. Such penalties enable using a standard \ac{gp} model over the unconstrained domain $\domain$, allowing for arbitrary acquisition procedures including \ac{gibo}. However, a fixed penalty effectively corresponds to fitting an extended objective function with a large discontinuity at the boundary between $\domain_S$ and $\domain_C$. Discontinuities are difficult to model with \acp{gp}, especially since we assume the function is differentiable (\cf~Assumption~\ref{ass:gp}). Instead, virtual data points are adaptive penalties that are chosen such that their value is not `too far' from the model predictions and, therefore, avoid large differences in function values while still guiding the optimization process away from observed crashes. Following \citet{stenger2022benchmark}, the adaptive penalty of the virtual observation is set to
\begin{equation}
    \hat y_i = \text{max}\left(\mu_\mathcal{D}(\hat \loc_i), \mu_\mathcal{D}(\loc_*) \right)  + \beta \sqrt{ k_\mathcal{D}(\hat \loc_i)},
\end{equation}
where the $\hat \loc_i$ are the observed crash locations, $\loc_*$ is the current parameterization, and $\beta > 0$ is a problem dependent parameter.
The virtual observations are added to the dataset, resulting in an augmented dataset $ \mathcal{\hat D} = \mathcal{D} \, \cup \, \{(\hat X, \hat y)\}$. 
The parameter $\beta$ determines the relative magnitude of the adaptive penalty with respect to the \ac{gp} model. Essentially, it encodes how unexpected a crash is for the model.
In general, determining a $\beta$ for the problem at hand can be difficult. However, setting $\beta = 3$ empirically worked well. It means that the `performance' of a crash is outside the $99\%$ confidence interval of the model.
Furthermore, we enforce a lower bound so the penalty is always larger than the posterior mean of current parameterization $\loc_*$. This ensures a non-negative slope between $\loc_*$ and all crash locations $\hat\Loc$, which can counteract cases where $f$ exhibits steep gradients towards the constraint. An example of the effect of virtual observations on the \ac{gp} posterior is shown in Fig.~\ref{fig:gp}.

\subsection{Gradient Information Bayesian Optimization under Crash Constraints}\label{ssec:cgibo}

\begin{algorithm}[t]
\small
\caption{\ac{crashgibo}}\label{algo:cgibo}
\begin{algorithmic}[1]
    \State \textbf{Input}: initial parameter $\loc_0$, batch sizes $b_k$, stepsizes $\eta_k$
    \State $\loc_* \leftarrow \loc_0$, $k=0$
    \State $y \leftarrow f(\loc_0) + \epsilon$, $\mathcal{D}  \leftarrow \{(\loc_0, y)\}$ \Comment{Optionally evaluate $\loc_0$}
    \Repeat
    \State $\Loc_{\textrm{DoE}} = \argmin_{X'} \alpha_{TV}(x_*, X')$ \Comment{Next batch (\cf\eqref{eq:gibo_doe})}
    \State $\bm{y} \leftarrow f(\Loc_{\textrm{DoE}}) + \epsilon$
    \State $\mathcal{D}  \leftarrow \mathcal{D} \cup \{(\Loc_{\textrm{DoE}}, \bm{y})\}$ \Comment{Update dataset}
    \State Build virtual dataset $ \mathcal{\hat D}$ (cf. Section~\ref{ssec:vdata})
    \State $\loc_* \leftarrow \loc_* - \eta_k \, \nabla\mu_\mathcal{\hat D}(\loc_k)$ \Comment{Gradient update}
    \State $y \leftarrow f(\loc_*) + \epsilon$ \Comment{Evaluate $x_*$  and update dataset}
    \If {$\loc_*$ is not feasible}
        \State $\loc_* \leftarrow \argmin_{x \in X} \mu_\mathcal{\hat D}(x)$ \Comment{Reset to a feasible $x$}
    \EndIf
    \State Optimize \ac{gp} hyperparameters.
    \Until{$\left|\mathcal{\hat D}\right| \geq K$} 
    \State \Return $\loc_*$
\end{algorithmic}
\end{algorithm}

We introduce a modification of the \ac{gibo} algorithm \cite{mueller2021local} for crash-constrained optimization problems called \ac{crashgibo} in Algorithm~\ref{algo:cgibo}.
The modifications to the original \ac{gibo} algorithm include (i) batch evaluations, (ii) a virtual dataset for the crashed observations, and (iii) resetting to a feasible parameterization if an update fails.
The core idea of \ac{gibo} is to evaluate the parameters that reduce the uncertainty of the gradient at the current iterate $x_*$ and then use the gradient estimate of the \ac{gp} to perform gradient descent
\begin{equation*}\label{eq:gradstep}
    \loc_{*,k+1} = \loc_{*,k} - \eta_k \, \nabla\mu_\mathcal{D}(\loc_k),
\end{equation*}
where $k$ denotes the iterate of \ac{crashgibo}.
Before each gradient step, the function is evaluated at $b$ locations that minimize the total variance of $p(\nabla f(\loc_*))$ (Section~\ref{ssec:gi}), where $b$ is the batch size. Specifically, we implement \eqref{eq:gi} in an automatic differentiation library and use a standard optimizer to find the \ac{doe}.

Our proposed algorithm \ac{crashgibo} needs to deal with crashes during evaluation and updates. If an evaluation crashes, we utilize the virtual dataset introduced in Section~\ref{ssec:vdata} to build a posterior over the whole search domain $\domain$. The virtual data points discourage further exploration near the crashed locations. Additionally, the adaptive penalty biases the gradient away from these areas, as illustrated in Fig.~\ref{fig:gp}.
After each update, the algorithm evaluates the new parameterization to ensure it is feasible. If not, $x_*$ is reset to a known feasible location from past evaluations $\Loc$. Assuming deterministic crashes, resetting to a known feasible evaluation guarantees a viable solution after each improvement step.
Specifically, we reset the parameters to the past evaluation with the minimal posterior mean $\loc_* = \argmin_{x \in X} \mu_\mathcal{\hat D}(x)$.

In general, \ac{gibo} can be used with any twice differentiable covariance function. However, we opt for the Gaussian kernel for our experiments since controller tuning objective functions are often smooth.
The Gaussian kernel is defined as (\cf\cite{rasmussen2006gaussian})
\begin{equation*}
    k_G(\bm{x}, \bm{x'}) = \sigma_f \exp{\left( \frac{1}{2} (\bm{x}-\bm{x'})^\top L^{-2} (\bm{x}-\bm{x'})\right)},
\end{equation*}
where $\sigma_f$ is the signal variance and $L$ is the positive-definite lengthscale matrix.
Using this kernel allows us to automatically adapt the stepsize $\eta_k$ of the gradient update to the lengthscale of the \ac{gp}
\begin{equation}\label{eq:stepsize}
    \eta_k = \frac{\hat\eta_k}{\sqrt{\nabla\mu_{\mathcal{D}}(\loc_*)^\top L^{-2} \nabla\mu_{\mathcal{D}}(\loc_*)}},
\end{equation}
where $\hat\eta_k$ is a chosen step size. Equation \eqref{eq:stepsize} scales the stepsize based on the expected correlations of function values. See \citet{mueller2021local} for more details in gradient normalization with \ac{gp} lengthscales. 

\section{Simulation Results}

This section provides an overview of the coupled tank system used to assess the performance of \ac{crashgibo} in controller tuning tasks. We also introduce the control algorithms under consideration and present the results of the simulated controller tuning process using \ac{crashgibo}.

\subsection{Process Description}
\label{ssec:process_desc}

\begin{figure}[t]
\vspace{2mm}
\centering
\includegraphics[width=0.48\textwidth]{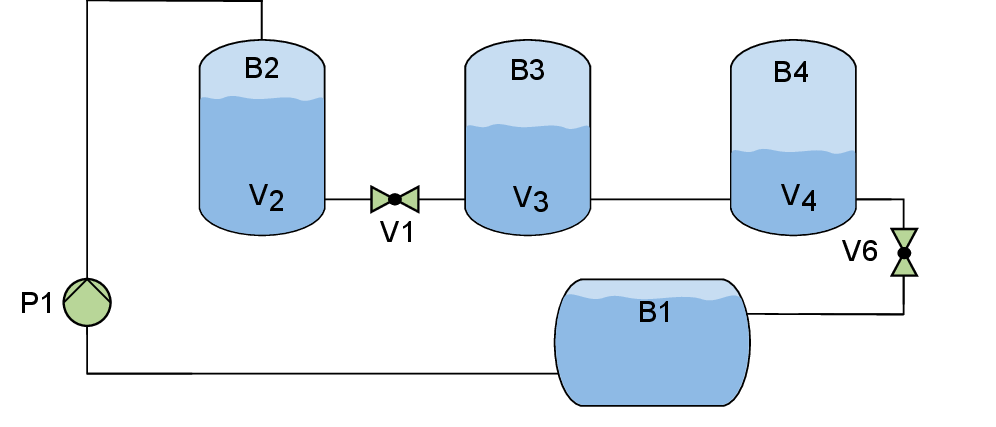}
\caption{Diagram of the coupled tank system, with controllable pump and valves.}
\label{fig:tank_schematic}
\end{figure}

The subject of examination in this study is a coupled tank system, illustrated in Fig.~\ref{fig:tank_schematic}. The system consists of three tanks, labeled $\text{B}2$, $\text{B}3$, $\text{B}4$  as well as one reservoir tank, B1. The pump $\text{P}1$, and the valves $\text{V}1$ and $\text{V}6$ can be actuated by the controller. The objective of the control is formulated in terms of the water levels in the three tanks, $\text{V}_2$, $\text{V}_3$ and $\text{V}_4$ respectively.
We formulate a non-linear state space representation of the system as 
\begin{equation}\label{eq:state_space}
\begin{aligned}
\dot{\state} &=\zeta(\mathbf{\state, \action}),
\end{aligned}
\end{equation}
where $\zeta$ is the dynamics, $\mathbf{\state} = \begin{bmatrix}
    \text{V}_{2} & \text{V}_{3} & \text{V}_{4}
\end{bmatrix}^{\top}$ are the states and 
$\action = \begin{bmatrix}
    \text{U}_{\text{P}} &  \text{U}_{\text{V}1}  & \text{U}_{\text{V}6}
\end{bmatrix}^{\top}$ are the inputs of the system. 
The coupled tank system used here is described in more detail in \citet{Scheurenberg23}. Despite careful modeling and empirical validation, there are discrepancies between the model and the dynamics.
For the control algorithms we linearize the model at the operating point $\text{P}$ with $\state_{\text{P}}=\begin{bmatrix}
    8 & 6 & 5
\end{bmatrix}\times10^{-3}$ and $\action_{\text{P}}=\begin{bmatrix}
    70.7 & 43.3 & 44.7
\end{bmatrix}$ and discretized with $10 \, \text{Hz}$  (\cf\cite{Scheurenberg23}).

\subsection{Process Control}
\label{ssec:process_control}
Three common controller configurations are employed to assess the automatic tuning algorithm \ac{crashgibo}: a cascaded control system using PI controllers, as well as an \ac{lqi} controller and a linear \ac{mpc} scheme. In the following, we will briefly overview the control loop structures. Additionally, we introduce the respective tuning parameters and objective functions of the five optimization problems summarized in Tab.~\ref{tab:simcases}.

The cascaded control comprises two distinct feedback loops. The outer loop regulates the water level $\text{V}_4$ by modifying the volumetric flow entering tank $\text{B}2$. The inner loop regulates this volumetric flow entering tank $\text{B}2$, by using $\text{U}_{\text{P}}$ as the only controlled variable. The volumetric flow generated by the pump is accessible as a measured variable within this control system. Here, we consider two cases. In the first case, we tune only the outer loop: $\loc = \left[ k_\mathrm{p,out}, k_\mathrm{i,out} \right]$. In the second case, all parameters for the cascaded structure are optimized. Furthermore, the position of valve $\text{V}1$ is a tuning variable but is kept constant over one episode: $\loc = \left[ k_\mathrm{p,out}, k_\mathrm{i,out}, k_\mathrm{p,in}, k_\mathrm{i,in}, \text{U}_{\text{V}1}  \right]$.  

Both the \ac{lqi} and the \ac{mpc} are multiple-input multiple-output (MIMO) control algorithms. The controlled variables are the water levels $\text{V}_{2}$,$ \text{V}_{3}$,$\text{V}_{4}$ and the manipulated variables are $\text{U}_{\text{P}}$, $ \text{U}_{\text{V}1}$, $\text{U}_{\text{V}6}$. The \ac{mpc} and the \ac{lqi} both use a linear model obtained by linearizing \eqref{eq:state_space} at a fixed operating point (\cf Section~\ref{ssec:process_desc}).

Since the process is non-linear, a linear \ac{mpc} is not guaranteed to achieve zero steady-state error. An extended Kalman filter (EKF), used as a disturbance estimator, addresses this issue. The EKF and the \ac{mpc} have several tuning parameters, such as the weighting matrices and the \ac{mpc} control and prediction horizons. In principle, \ac{bo} can also optimize  the horizons \cite{STENGER2020}, but here they are fixed to the values in \citet{Scheurenberg23}. For this study, we optimize the entries of the the weighting matrix $\mathbf{Q}_{\mathrm{MPC}} = \mathrm{diag} \left[10^{\locentry_{1}}, 10^{\locentry_{2}}, 10^{\locentry_{3}}\right]$ penalizing the tracking error. Additional tuning parameters are the entries corresponding to the disturbance process noise of the EKF $\mathbf{Q}_{\mathrm{EKF}} = \mathrm{diag} \left[ 1, 1, 1, 10^{\locentry_{4}}, 10^{\locentry_{5}}, 10^{\locentry_{6}}\right]$, resulting in a total of 6 optimization variables.
This parameterization enables tuning the weighting matrices over many orders of magnitudes and, empirically, yields well tuned controllers \cite{STENGER2020}.

Similarly to an \ac{mpc}, a standard LQR is not guaranteed to achieve zero steady-state error for a non-linear process. Therefore, using \ac{lqi} adds integral error states to the state vector. For the \ac{lqi}, all diagonal entries of the weighting matrices are optimized: $\mathbf{R}_{\mathrm{LQI}} = \mathrm{diag} \left[ 1, 10^{\locentry_{1}}, 10^{\locentry_{2}}\right]$, $\mathbf{Q}_{\mathrm{LQI}} = \mathrm{diag} \left[ 10^{\locentry_{3}}, \dots , 10^{\locentry_{8}} \right]$.     

In both PI-tuning cases, one episode consists of one step of the reference for $\text{V}_{4}$. The objective function is the root mean squared tracking error (RMSE) 
\begin{equation*}\label{eq:rmse}
    f_{\mathrm{RMSE}}(\loc) = \mathbb{E} \left[\sqrt{\frac{1}{T} \int_0^{T} (\text{V}_{4}(t,\loc) - \text{V}_{4,\mathrm{ref}}(t))^2 \mathrm{d}t }\right],
\end{equation*}
where $T$ is the episode length. In cases where the controller is not used for a short episode but for continuous operation, a representative episode with typical disturbances must be defined by domain experts. This is a general problem for all data-based tuning methods. A typical scenario is the closed-loop step response. 
Here, the objective is evaluated using a single experiment: a noisy approximation of the true expectation.
In the \ac{mpc} and \ac{lqi} cases, the objective function and the reference trajectory are chosen differently to highlight the broad applicability of \ac{bo}. Here, the weighted sum of the mean absolute tracking error (MAE) of each tank is minimized
\begin{equation*}
   f_{\mathrm{MAE}}(\loc) = 0.5 f_{\mathrm{V},2}(\loc) + 0.25 f_{\mathrm{V},3}(\loc)+ 0.25 f_{\mathrm{V},4}(\loc),
\end{equation*}
where 
\begin{equation*}
    f_{\mathrm{V},i}(\loc) = \mathbb{E} \left[ \frac{1}{T} \int_0^{T} | \text{V}_{i}(t,\loc) - \text{V}_{i,\mathrm{ref}}(t) |\mathrm{d}t \right].
\end{equation*}
Instead of one reference step, two consecutive reference steps for all water volumes are evaluated.

A simulation is aborted, \ie a crash occurs, in case the first tank exceeds a critical volume. In practice, this prevents a tank from overflowing. For the PI-case, this crash $V_2$ is set to $7 \, \mathrm{l}$ and $8 \, \mathrm{l}$, respectively. In the cascaded PI case, $8 \, \mathrm{l}$  and for the remaining cases, $7.5 \, \mathrm{l}$ is chosen.

\begin{table} [t]
\centering
\caption{Summary of the simulation test cases.}
\begin{tabular}{rrrr}
Case Name  & No. of Params. & Objective & Crash $V_2$\\
 \midrule
PI ($V_2 < 8 \, \mathrm{l}$) & 2 & RMSE &  $8 \, \mathrm{l}$\\
PI ($V_2 < 7 \, \mathrm{l}$) & 2 & RMSE & $7 \, \mathrm{l}$\\
cascaded PI      & 5 & RMSE & $8 \, \mathrm{l}$\\
MPC + EKF        & 6 &  MAE & $7.5 \, \mathrm{l}$\\
LQI              & 8 & MAE  & $7.5 \, \mathrm{l}$\\

\end{tabular}

\label{tab:simcases}
\end{table}

\subsection{Tuning Results}\label{subsec:sim_results}

\begin{table} [t]
\caption{\ac{crashgibo} hyperparameters for all simulations and experiments.}
\centering
\begin{tabular}{rr}
Hyperparamter& Value\\
 \midrule
 $b_k$& $d+1$\\
$\hat\eta_k$& $0.25 \dots 0.125$ (with cosine decay)\\
$L$& 0.25 $I$\\
 $\sigma_f$&$0.5$\\
$\mu(x)$& $1$\\
\end{tabular}
\label{tab:hyperparams}
\end{table}

\begin{figure*}
    \centering
    \input{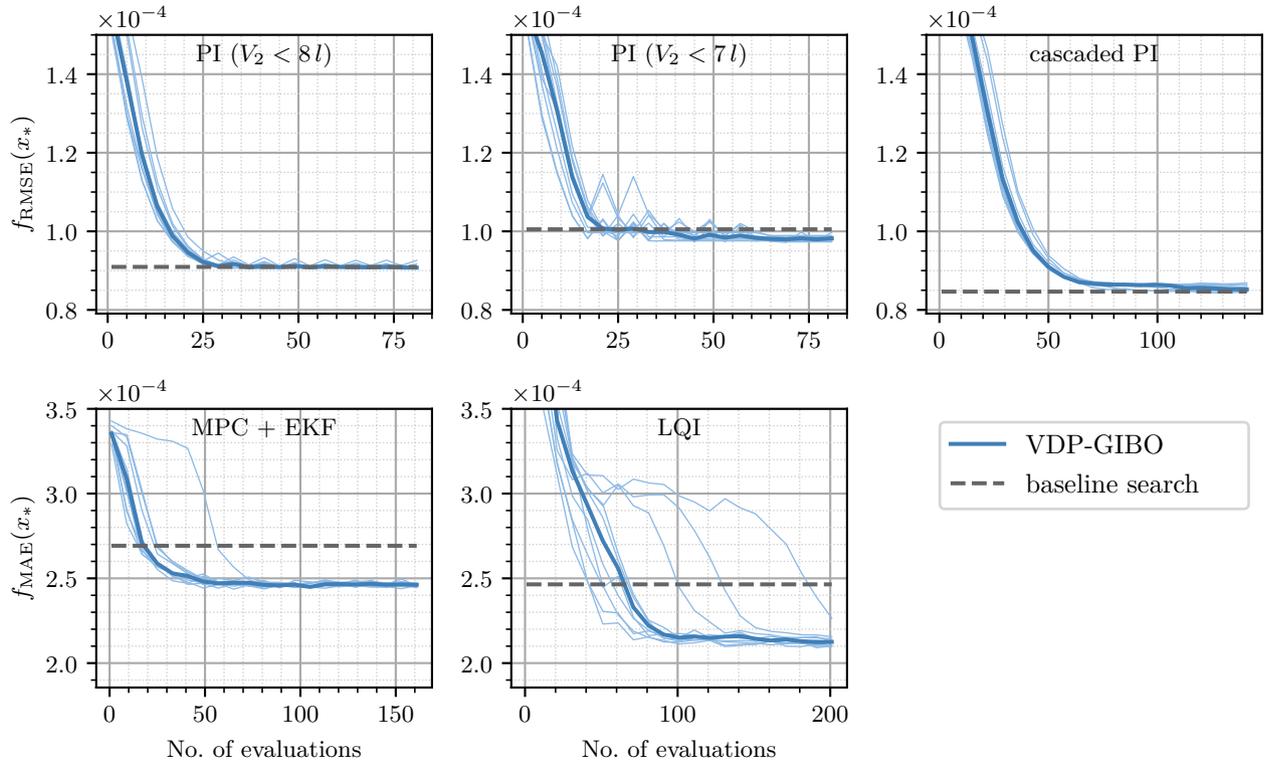}
    \caption{\textbf{Simulation results on crash constrained controller tuning problems:}
    \ac{crashgibo} is able to solve $2$-(PI),4-(cascaded PI),6-(\ac{mpc}) and 8-dimensional (\ac{lqi}) controller tuning problems in a handful of evaluations. The controller performance shown as the median over $10$ runs with randomized initial controller parameters. The individual runs are shown as thin lines and demonstrate the low variability in tuning results with the proposed method.
    As baseline (dashed line), we draw parameters uniformly at random from the search domain and chose the best evaluation. The number of evaluation is the same as for \ac{crashgibo}.
    Please note that the objective functions are different between the PI and the MIMO (\ac{lqi} and \ac{mpc}) controllers.}
    \label{fig:sim_results}
\end{figure*}
\begin{figure*}
    \centering
    \input{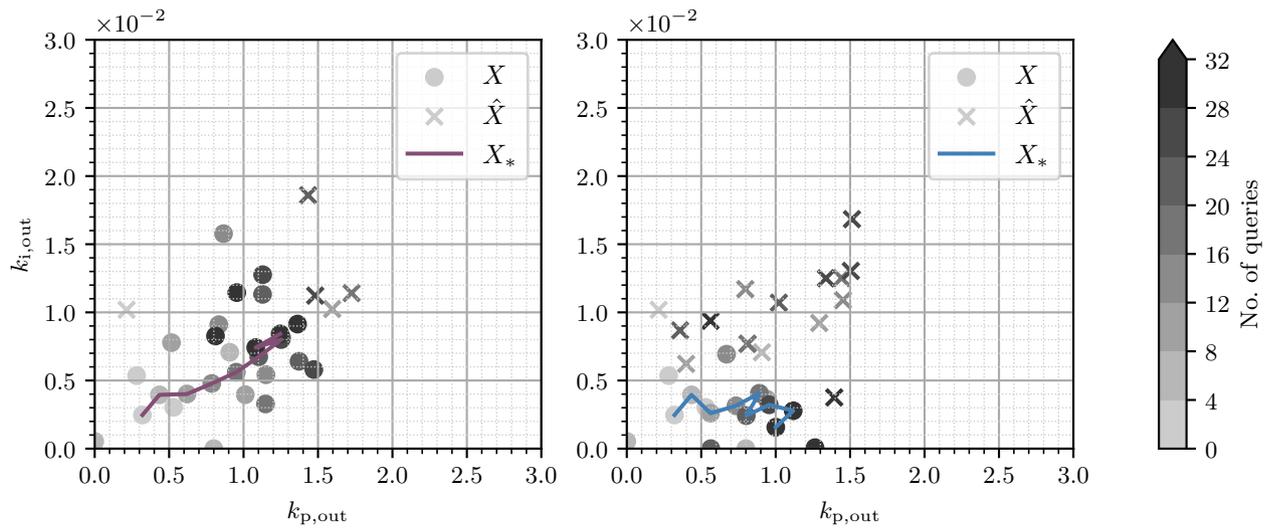}
    \caption{\textbf{Evaluations in the parameter space} for PI ($V_2 < 8 \, \mathrm{l}$) (left) and PI ($V_2 < 7 \, \mathrm{l}$) (right). We show the first eight improvement steps and the corresponding evaluation locations $\Loc$, improvement steps $\Loc_*$ and crashes $\hat \Loc$. Due to the tighter constraints on the right, the feasible optima changes. The virtual observations change the gradient such that the algorithm can estimate this new local optimum. The majority of the parameter space remains unexplored, increasing data-efficiency.}
    \label{fig:sim_path}
\end{figure*}

We evaluate the performance of \ac{crashgibo} in the five settings described in the previous section. Each control algorithm is tuned ten times with randomized initial controller parameterization and different noise realizations. The initial controller parameterization is chosen from a small set in the feasible region $\domain_S$ where the performance is relatively low.
We use the same hyperparameters in all experiments (Tab.~\ref{tab:hyperparams}).
These parameters were obtained manually from initial experimentation with the PI controller.
Since the hyperparameters were not tuned to the specific problems and were chosen based on basic knowledge of the problem domain, we conclude that the controller tuning with \ac{crashgibo} is not very sensitive to the choice of hyperparameters.

We compare the tuning result of \ac{crashgibo} with the result of random search, where we draw controller parameterization uniformly from the search domain $\domain$, evaluate them and choose the best one. We use the same number of evaluations for the random search baseline as for \ac{crashgibo}.
The optimization results are shown in Fig.~\ref{fig:sim_results}.
The PI tuning problems (Fig.~\ref{fig:sim_results}, top) are relatively easy and random search can solve them within a few evaluations. Our method recovers similar solutions within the given budget and even finds slightly better solutions for the more constrained problem (Fig.~\ref{fig:sim_results}, top-middle). However, for these easy tuning problems, \ac{crashgibo} usually takes more evaluations than random search. The reason is that the initial guess is purposefully poor, and \ac{crashgibo} performs a local search, requiring a few steps before leaving this high-cost initial region. For the more complex and higher dimensional tuning problems (Fig.~\ref{fig:sim_results}, bottom), random search is not a viable tuning strategy, and \ac{crashgibo} finds controllers with significantly better performance within the same evaluation budget, highlighting the data-efficiency of our proposed method. 
Since the difficulty of a problem is unknown a priori, \ac{crashgibo} yields more consistent results.

The resulting control performances are consistent with control engineering intuition: Tightening the upper bound for V$2$ results in a worse tracking behavior. Additionally, adding degrees of freedom to the tuning task by adjusting the inner control loop and the first valve position can increase closed-loop performance.
For the LQI tuning problem, we allow for the relative weighting between the different penalties, $R_{\mathrm{LQI}}$, on the control input. In contrast, for the MPC, the weighting matrix for the input is fixed. Generally, control engineering expertise is required to choose influential tuning parameters for a given controller structure. Additionally, it emphasizes the importance of \ac{crashgibo} being able to address high-dimensional tuning problems. 

For the two-dimensional tuning problems, we show the evaluations of \ac{crashgibo} in the parameter space in Fig.~\ref{fig:sim_path}. The algorithm is able to find a local optimum without exploring the whole search space $\domain$, leading to a data-efficient tuning process and relatively few crashes, even when a large part of the space is infeasible.
For the problem with tighter constraints on $V_2$ a larger part of the search space, including the optimum, is infeasible. The virtual data points enable a \ac{gp} model over the whole search space, and \ac{crashgibo} converges to a feasible estimate. 

\section{Experimental Results}

\begin{figure}[t]
    \centering
    \includegraphics[width=0.75\columnwidth]{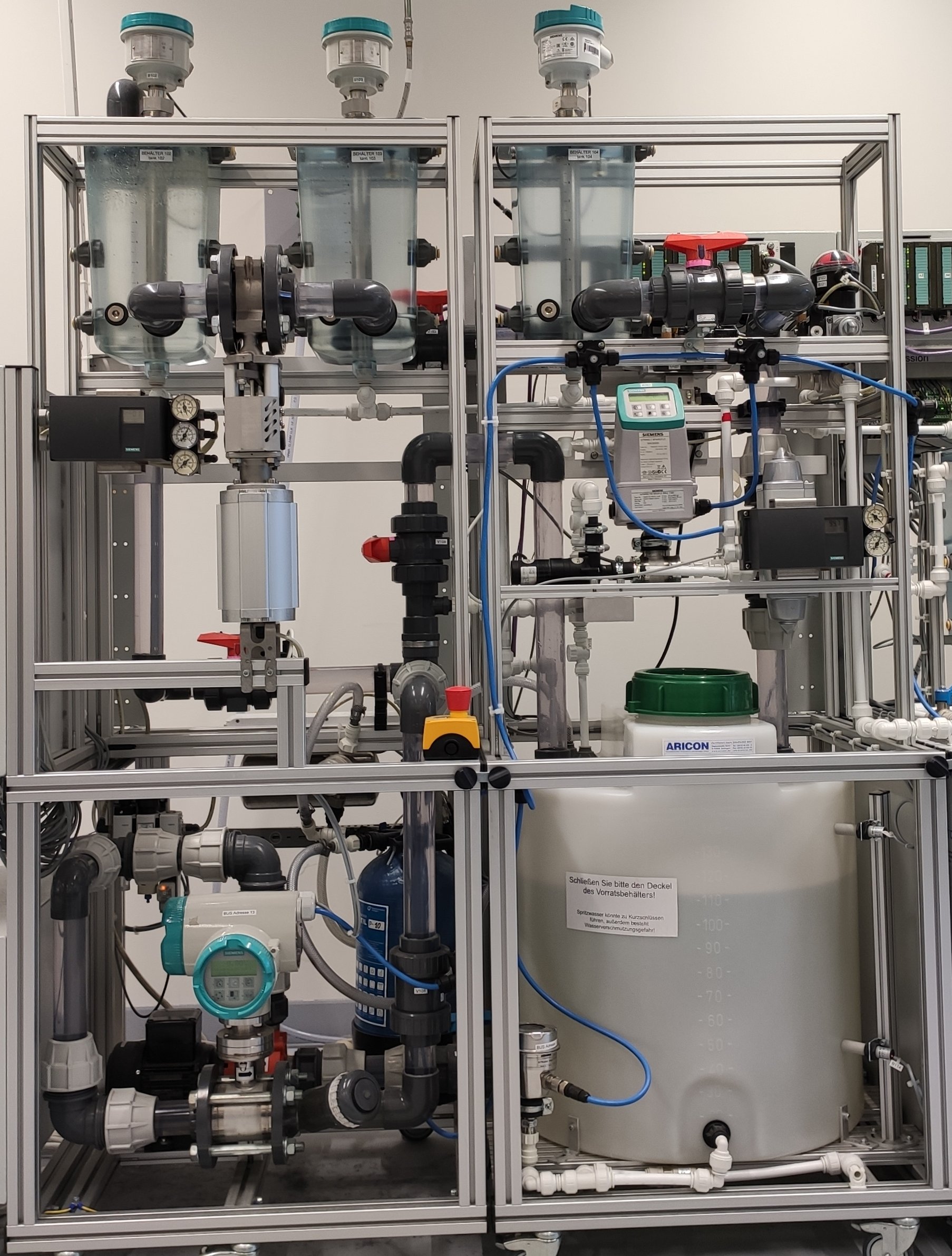}
    \caption{\textbf{Experimental setup:} The top three tanks correspond to tanks B1, B2, and B3 (from left to right). The black boxes in the middle of the image are the pneumatic valves V1 and V6 (from left to right). At the bottom of the picture are the pump and the reservoir tank B1.}
    \label{fig:tank}
\end{figure}

We repeated the tuning experiments for the PI and the cascaded PI controller on a hardware test bed that implements the process described in Section~\ref{ssec:process_desc}. The testbed is part of the model factory at the \emph{Institute of Automatic Control, RWTH Aachen University} and is depicted in Fig.~\ref{fig:tank}.
Analogous to the description of the simulation model, water is pumped into tank B2 by pump P1. The volumetric flow generated by the pump as a function of the control input is measured using a flow sensor. The pneumatic valve V1 controls the volume flow from B2 to B3. Subsequently, tank B3 is connected to tank B4 without a controllable valve. Tank B4 is connected to the reservoir tank B1 via V$6$. Each tank has a level sensor.
All hyperparameters of \ac{crashgibo} are the same as in Section~\ref{subsec:sim_results}.

\begin{figure}[t]
    \centering
    \input{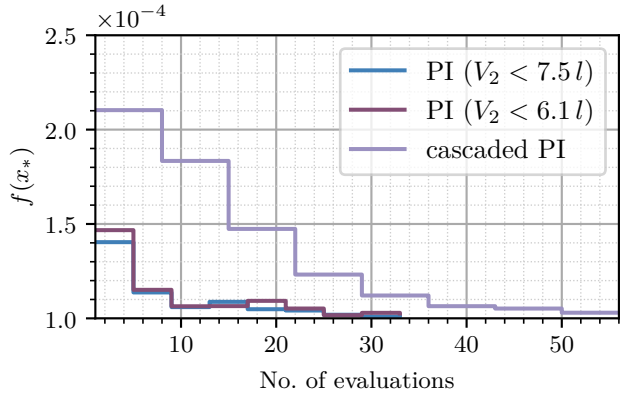}
    \caption{\textbf{Experimental results for crash-constrained PI control:} Control performance of the PI controller during the tuning process. Despite the crash constraints, \ac{crashgibo} is able to use the data efficiently and improve by approx. $33\%$ for the PI controller and by $50\%$ for the cascaded PI controller.}
    \label{fig:result_hardware}
\end{figure}

\begin{figure}[t]
    \centering
    \input{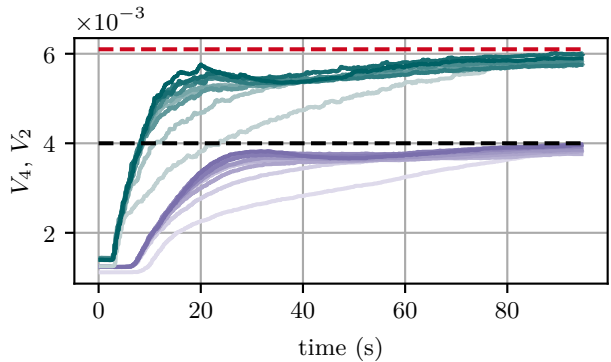}
    \caption{\textbf{Time domain of the PI controller:} Step response of $V_4$ (purple) and $V_2$ (petrol) with the PI controller. After tuning, the state tracks the desired reference (black) significantly better. Darker colors indicate the behavior later in the tuning process. The state $V_2$ stays below the constraint (red).}
    \label{fig:time}
\end{figure}

For the first two experiments, we tuned the outer control loop of the cascaded PI controller: $ \loc = \left[ k_\mathrm{p,out}, k_\mathrm{i,out} \right]$. The inner controller was set to hand-tuned values. In the first experiment, we set the crash constraint on the maximum water level in all tanks to $7.5 \, \text{l}$. When the water level reaches this level, the experiment is aborted. 
In this setting, most controller parameterizations are feasible, and due to the local exploration of \ac{crashgibo}, all evaluations during tuning were feasible.
We ran the experiment for eight iterations, corresponding to $33$ evaluations or approximately $1.5$ hours. The controller performance improved by ca.\ $33\%$ (Fig.~\ref{fig:result_hardware}). Most of the improvement was achieved in the first three gradient steps, highlighting the efficiency of local \ac{bo} for controller tuning.

The second experiment is run with a maximum water level of $6.1 \, \text{l}$. With this constraint, many parameters lead to an emergency system stop, preventing the experiment from completion. 
In the tuning experiment, three controller evaluations crashed, with two crashes during the exploration phase and one during an update. Our proposed algorithm \ac{crashgibo} is still able to achieve a similar tuning result as in the previous setting despite the more difficult tuning task (Fig.~\ref{fig:result_hardware}). The time domain behavior of the controlled and constrained state is depicted in Fig.~\ref{fig:time}.

In the third experiment, we jointly tune the cascaded PI controller and the position of valve $V1$: $\loc = \left[ k_\mathrm{p,out}, k_\mathrm{i,out}, k_\mathrm{p,in}, k_\mathrm{i,in}, \text{U}_{\text{V}1}  \right]$.
The initial parameters are chosen such that the initial performance is poor. Due to the higher dimensional search space,  eight iterations of \ac{crashgibo} require a budget of $54$ evaluations or $2.7$ hours. The algorithm improves the control performance by approximately $50\%$ and, due to its local exploration behavior, never leaves the feasible region.

\section{Conclusion}

In this article, we propose \ac{crashgibo}, a novel controller tuning algorithm for optimization problems under crash constraints, and demonstrate its data-efficiency on three standard control algorithms: PI control, \ac{lqi}, and \ac{mpc}.
While the proposed algorithm has a set of hyperparameters, these were always the same for all the results presented in the paper. This points towards the applicability of the algorithm for general controller tuning problems. However, other tuning problems may require additional effort and evaluations to find suitable parameters.

Sample-efficient and intuitive controller tuning, especially for well-proven policy/controller structures such as PI, \ac{lqi}, and \ac{mpc} using \ac{crashgibo}, can lead to overall higher control performance by combining control engineering expertise with data-driven techniques. Control engineering expertise is required to choose a suitable controller structure and formulate the tuning task by defining the objective function and tuning parameters. The data-driven optimization explores the search space locally, resulting in -- compared to global exploration -- gradual changes in time domain behavior. This enables a more intuitive understanding of the tuning progress.

Automated tuning can help control engineers to fairly compare the applicability of different controller structures to practical control engineering tasks. Tuning structures with respect to identical objective functions eliminates the influence of hand-picked parameter values on the comparison. The applicability of \ac{crashgibo} to higher dimensional problems enables the control engineer to design efficient controller structures with many parameters.   

This paper assumes that a mechanism to detect a crash and reset the system to its initial state is present. Such resets might be challenging for fully automated controller tuning and require a backup controller to take over in case of failures (\eg\cite{baumann2021gosafe,schurmann2022formal}).

\begin{acknowledgement}
The authors thank P. Brunzema, J. Menn, and F. Solowjow for constructive criticism of the manuscript.
\end{acknowledgement}

\addtolength{\textheight}{-4cm}

\bibliographystyle{unsrtnat}
\bibliography{refs}

\end{document}